\documentclass[preprint,nofootinbib,superscriptaddress]{revtex4}

\usepackage{epsfig}
\usepackage{subfigure}
\usepackage{amssymb}
\usepackage{amsmath}
\def\beq{\begin{equation}}
\def\eeq#1{\label{#1}\end{equation}}
\def\eeqn{\end{equation}}
\def\beqa{\begin{eqnarray}}
\def\eeqa#1{\label{#1}\end{eqnarray}}
\def\eeqan{\end{eqnarray}}

\newcommand\one{\leavevmode\hbox{\small1\normalsize\kern-.33em1}}

\begin{document}
 
\preprint{
 {\vbox{
 \hbox{\bf MADPH-07-1490}
 \hbox{}
 \hbox{}
 }}}
\title{$\cal T$-Anomaly Induced LHC Signals}

\author{V. Barger$^{1}$, Y. Gao$^{1}$, and Wai-Yee Keung$^{2}$\\[2ex]
\small\it $^1$Department of Physics, University of Wisconsin, 
Madison, WI 53706\\
\small\it $^2$Department of Physics, University of Illinois at Chicago, 
IL 60607-7059}
\noaffiliation

\begin{abstract}
$\cal T$-parity in the Little Higgs model could be violated by
anomalies that allow the lightest $\cal T$-odd $A_H$ to decay into
$ZZ$ and $W^+W^-$.  We analyze these anomaly induced decays and
the two-particle and
the three-particle decay modes of other heavy quarks and bosons in
this model which yield unique Large Hadron Collider (LHC) signals with
fully reconstructable events. $\cal T$-odd
quarks in the Little Higgs model are nearly degenerate in mass and
they decay by almost identical processes; however, members of the
heavy Higgs triplet follow distinct decay modes.  The branching
fractions of three-body decays increase with the global symmetry-breaking
energy scale $f$ and are found to be at the level of a few percent in heavy quark decays
while they can reach up to 10\% for heavy bosons.

\end{abstract}

\maketitle

The Higgs mass in the Standard Model (SM) receives large radiative
corrections from the short-distance physics at the cutoff
scale. Fine-tuning in the Higgs sector becomes an eminent problem,
especially when the SM predictions are confronted with precision
electroweak data.\cite{Barbieri:2000gf} In order to naturally
alleviate the quadratic divergent contributions, new particles are
expected to exist with TeV scale masses.

The Little Higgs mechanism\cite{LittleHiggs} makes use of the light
mass property of the pseudo-Nambu--Goldstone boson (pNGB) to protect the
Higgs mass from the one-loop quadratic divergence: its mass receives
one-loop radiative corrections from the new TeV scale particles, which
cancel the corrections from Standard Model fermion and boson loops.


\section*{Little Higgs with $\cal T$ Parity}

One of the simplest implementations of such a mechanism is the
Littlest Higgs (LH) model \cite{Arkani-Hamed:2002qy} based on
$$ G=SU(5)  \quad \hbox{and}  \quad 
G_1\otimes G_2 =[SU(2)_1\otimes U(1)_1]\otimes[SU(2)_2\otimes U(1)_2 ]
\ .$$
At $f\sim 1$ TeV the initial $SU(5)$ global symmetry spontaneously
breaks down to an $SO(5)$ subgroup in the direction
\[ \Sigma_0=\left( \begin{array}{ccc}
 & & \one  \\
   & 1&    \\
 \one&  &   \\
\end{array} \right) \] 
where $\one$ is the identity matrix. 
After symmetry breaking at the energy scale $f$,
the dynamics near $\Sigma_0$ is described by the non-linear sigma field
$\Sigma=e^{\frac{2i}{f}X_a t_a} \Sigma_0$, where $t_a$ are the
pseudo-Nambu-Goldstone bosons (pNGB) 
associated with the 14 generators $X_a$ of the broken symmetry. An
$[SU(2)\times U(1)]^2$ subgroup of the $SU(5)$ is weakly gauged. 
Gauging each of the two $SU(2)\times U(1)$ leaves a different $SU(3)$ subgroup unbroken,
i.e. unless both $SU(2)\times U(1)$ are broken there will be a preserved $SU(3)$ symmetry and 
the Higgs field will be an exact massless Nambu-Goldstone field. Thus any loop contribution
to Higgs mass must involve couplings from both copies of $SU(2)\times SU(1)$. At
one loop level the leading contribution is only logarithmically divergent under this requirement.
This mechanism that protects the Higgs mass from quadratic divergneces is often referred as
"collective symmetry breaking".


$\Sigma_0$ breaks the full gauge group to the diagonal SM electroweak
group $SU(2)\times U(1)$ at energy scale $f$. Four pNGB's give TeV scale masses
to $W_H^\pm, W_H^3$ and $B_H$. $W_H^3$ and $B_H$
mix and form mass eigenstates $A_H$ and $Z_H$ in analog to the SM
photon and $Z$ boson. Other pNGB fields group into a
doublet identified as the SM Higgs and a weak triplet $\Phi$
\begin{equation}
\Pi=
\left(
\begin{array}{ccc}
{\bf 0}_{} & \frac{H}{\sqrt{2}} &  \Phi  \\
\frac{H^\dagger}{\sqrt{2}} & 0 & \frac{H^{\rm T}}{\sqrt{2}} \\
\Phi^\dagger & \frac{H^*}{\sqrt{2}} & {\bf 0}_{}
\end{array}
\right) \ ,
\hbox{ with } \ \ 
H=\left(
\begin{array}{c}
-i\pi^+\\
\frac{h+i\pi^0}{\sqrt{2}}
\end{array}
\right) \ ,
\Phi=\left(
\begin{array}{cc}
-i\phi^{++} & -i\frac{\phi^+}{\sqrt{2}}\\
-i\frac{\phi^+}{\sqrt{2}} & -i \frac{\phi^0+i\phi^p}{\sqrt{2}} 
\end{array}
\right)  ,
\end{equation}
where $\pi^+$ and $\pi^0$ in the Higgs doublet are eaten by SM weak
bosons.\cite{Han:2003wu} All the new particles are massive.
The new TeV scale particles $W^{\pm}_H, Z_H, A_H, \phi^{\pm}, \phi^{\pm\pm}, \phi^0, \phi^p$
couple to the Higgs field and 
cancel the quadratic radiative corrections to the
Higgs mass arising from their SM counterparts. 
 
However, the tight constraints from precision electroweak data
disfavor the LH model at a natural symmetry breaking scale $f\sim$1 TeV.
Phenomenological constraints on LH parameters\cite{EWC} push
the lower boundary of new physics up to about 10 TeV, but the
naturalness principle sets all dimensionless couplings to $\sim1$ and
requires the energy scale to be around 1 TeV. Thus the LH model needs an
energy scale higher than the natural value to stay consistent with
electroweak results. This tension is often referred to as the `little
hierarchy' problem. To address this issue, Cheng and Low proposed that
an additional discrete\cite{Cheng:2003ju,Low:2004xc} $\cal T$-Parity
can be imposed to relax\cite{Hubisz} the confrontation between theory
and experimental constraints.  

Similar to the matter parity in
supersymmetry, $\cal T$-parity is introduced as a global discrete
parameter.  It exchanges $[SU(2)_1\times U(1)_1]$ and $[SU(2)_2\times
U(1)_2]$. $\Sigma$ transforms under $\cal T$-parity as $\Sigma\rightarrow
\tilde{\Sigma}=\Sigma_0 \Omega \Sigma^\dagger\Omega \Sigma_0$ with
$\Omega ={\rm diag}(1,1,-1,1,1)$.

All SM particles and the LH heavy top quark $T_+$ are assigned $\cal T$-even in 
the Little Higgs model with $\cal T$-parity (LHT).
All other heavy particles are assigned $\cal T$-odd.  In the fermion sector, each SM
fermion is extended into a pair of SU(2) doublets $q_1$ and $q_2$ that
transform under $SU(2)_1$ and $SU(2)_2$. $\cal T$-parity interchanges
$q_1$ and $q_2$. Their $\cal T$-even combination is associated with
the SM fermion, while the $\cal T$-odd combination is the heavy
partner to the SM particle. Interaction terms $ -\kappa f (\bar{\Psi}_2 e^{i\Pi} \Psi_c
+\bar{\Psi}_1 \Sigma_0 \Omega e^{-i\Pi} \Omega\Psi_c)+{\rm h.c.}$ give
mass to $\cal T$-odd fermions
\begin{eqnarray}
M_{d_-}&\simeq& \sqrt{2}\kappa f,~~M_{u_-}\simeq \sqrt{2}\kappa f
\left(1-\frac{v_{SM}^2}{8f^2}+\cdots\right) \ ,
\label{Todd_mass}
\end{eqnarray}
where $v_{SM} = 246$ GeV is the Higgs vacuum expectation value in
the SM and $\kappa$ is a free parameter. For illustration, we take
$\kappa=1$ throughout this paper. In the interaction term $\Psi$ is
the SU(2) fermion doublet embedded into the SU(5) multiplet:
$\Psi_1=(q_1,0,{\bf 0}_2)^{\rm T}$, $\Psi_2=({\bf 0}_2,0,q_2)^{\rm T}$
and $\Psi_c=(q_c,\chi_c,\tilde{q}_c)^{\rm T}$.
Details are given in Ref.\cite{Hubisz}.

There is a special treatment in the top sector besides the heavy $\cal
T$-odd weak {\it doublet} $(b_-,t_-)=(u_3,d_3)$ of the third
generation.  In LH the large top quark loop correction demands an
additional vector-like weak {\it singlet} $T_+$ quark to stabilize the
Higgs mass. It is assigned even $\cal T$-parity in LHT and its $\cal T$-odd
partner is introduced as another top-like heavy quark $T_-$.

In the bosonic sector, the doublet (triplet) Higgs $H$ ($\Phi$) is
even (odd) under ${\cal T}$-parity. The ${\cal T}$-even combinations
of the gauge fields are the SM $SU(2)_L$ gauge bosons $(W^a_\mu)$ and
$U(1)_Y$ hypercharge gauge boson $(B_\mu)$; the ${\cal T}$-odd
combinations are ${\cal T}$-parity partners($A_H$, $W_H^{\pm}$, $Z_H$) of the SM gauge
bosons. The masses are given as
\begin{eqnarray}
M_{A_H} &=&\frac{g'f}{\sqrt{5}}
\left[1-\frac{5v_{SM}^2}{8f^2}+\cdots\right],~
M_{Z_H} \simeq M_{W_H}=gf \left[1-\frac{v_{SM}^2}{8f^2}+\cdots\right] \ .
\end{eqnarray}
The lightest ${\cal T}$-odd particle is $A_H$, which could be
a dark matter candidate if ${\cal T}$-parity was
strictly conserved.

The new $\cal T$-odd bosons have masses around a few hundred GeV. The
new fermions have higher masses near 1 TeV.  A typical mass spectrum
of heavy particles in LHT is shown in Table \ref{tab:mass_spectrum}.

\begin{table}[ht]
\begin{center}
\begin{tabular}{|c|c|c|}
\hline
Particle&Mass (TeV)&$\cal T$ parity\\
\hline
  $A_H$&0.24&$-$ \\
 $Z_H~(W_H)$ & 0.97&$-$ \\
$\phi$ & 1.7&$-$ \\
$T_-$ & 1.5 &$-$ \\
$u_-,c_-,t_-, d_-,s_-,b_-$ &  2.1&$-$ \\
$T_+$&2.1&+\\
$e_H^-,\mu_H^-,\tau_H^-,\nu_{eH}, \nu_{\mu H}, \nu_{\tau H}$&2.1&$-$ \\
\hline
\end{tabular}
\end{center}
\caption{\footnotesize Characteristic masses of the heavy partners of
the SM particles.  Here we take the scale $f=$1.5 TeV, $\kappa=1$, and
the top quark and Higgs boson masses to be $175$ and $200$ GeV,
respectively. The dependences of particle masses on $f$ are plotted in
Fig.\ref{fig:MASS}.  }
\label{tab:mass_spectrum}
\end{table}

\begin{figure}[ht]
\begin{center}
 \includegraphics[scale=0.6]{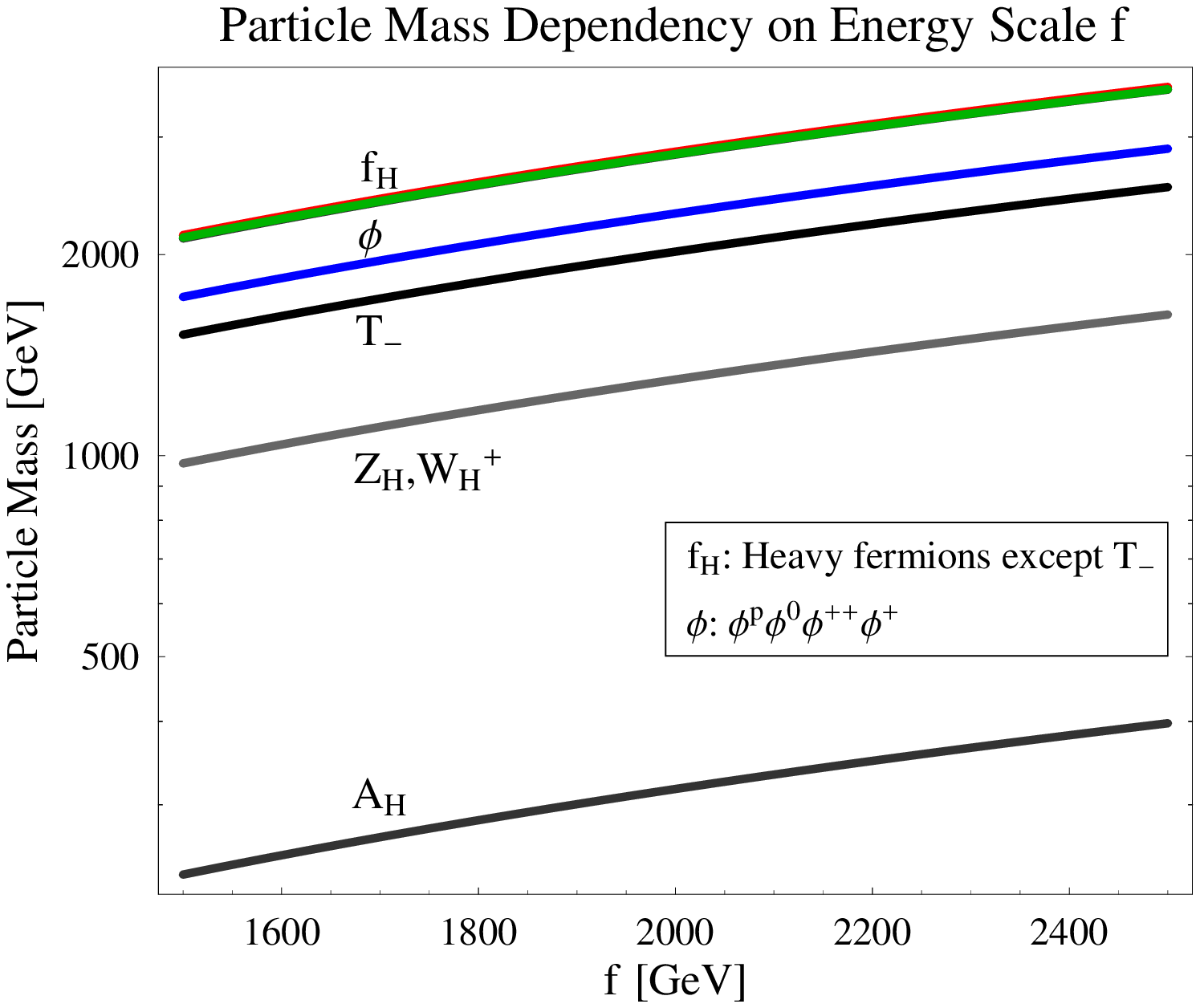}
\caption
{\footnotesize Masses dependency on symmetry breaking scale $f$ with $m_H$=200 GeV. Note
that most heavy fermions are very degenerate in mass. In this figure $f_H$ denotes the heavy 
fermions in the LHT model except for $T_-$, and $\phi$ denotes the heavy Higgs fields.}
\label{fig:MASS}
\end{center}
\end{figure}

The addition of $\cal T$-parity forbids $\cal T$-odd particles from mixing 
with their SM counterparts and leaves low-energy
observables unaffected by heavy particles at tree level. This
significantly loosens precision electroweak constraints on the
symmetry breaking scale $f$, permitting $f$ to be as low as 500 GeV at 
the expense of a high Higgs mass.\cite{Hubisz:2005tx} For
instance, $f=$ 1 TeV requires 280$ < m_h <$ 625 GeV.\cite{Casas:2005ev}

\section*{$\cal T$ Parity Violation in LHT}

A recent topological study by C. Hill and R. Hill finds that $\cal T$-parity
is violated by anomalies\cite{Hill:2007zv}, in which case the 4D
spacetime is a membrane embedded in a 5D bulk. The Little Higgs (LH)
lagrangian is reconstructed from a more general 5D bulk
lagrangian. The $\cal T$-parity plays a role similar to that of the
KK-mode parity: Symmetric under reflection in the 5th dimension, 
the zeroth mode of a 5D SU(3) gauge field is assigned $\cal T$-even and
identified with the vector field. The first mode of the 5D gauge field
transforms antisymmetrically under reflection in the 5th dimension,
and is identified with the $\cal T$-odd axial vector field.

In the Little Higgs model with $\cal T$-parity (LHT) 
pseudo-Nambu-Goldstone bosons introduce 
anomalous topological interactions at the global
symmetry breaking scale
$\sim\Lambda=4\pi f$. Consequently the Wess--Zumino--Witten (WZW)
term\cite{Wess:1971yu} that contains these topological effects must be
included into the full LHT Lagrangian and is essential for the UV
completion of the theory. Ref.\cite{Hill:2007zv} showed that $\cal T$-parity is
generally violated by anomaly; therefore the WZW term violates $\cal
T$-parity as well. 

\begin{figure}[t]
  \begin{center}
	\addtolength{\subfigcapskip}{-0.15\baselineskip}
	 \subfigure[$A_H$ decay]{\includegraphics[scale=0.60]{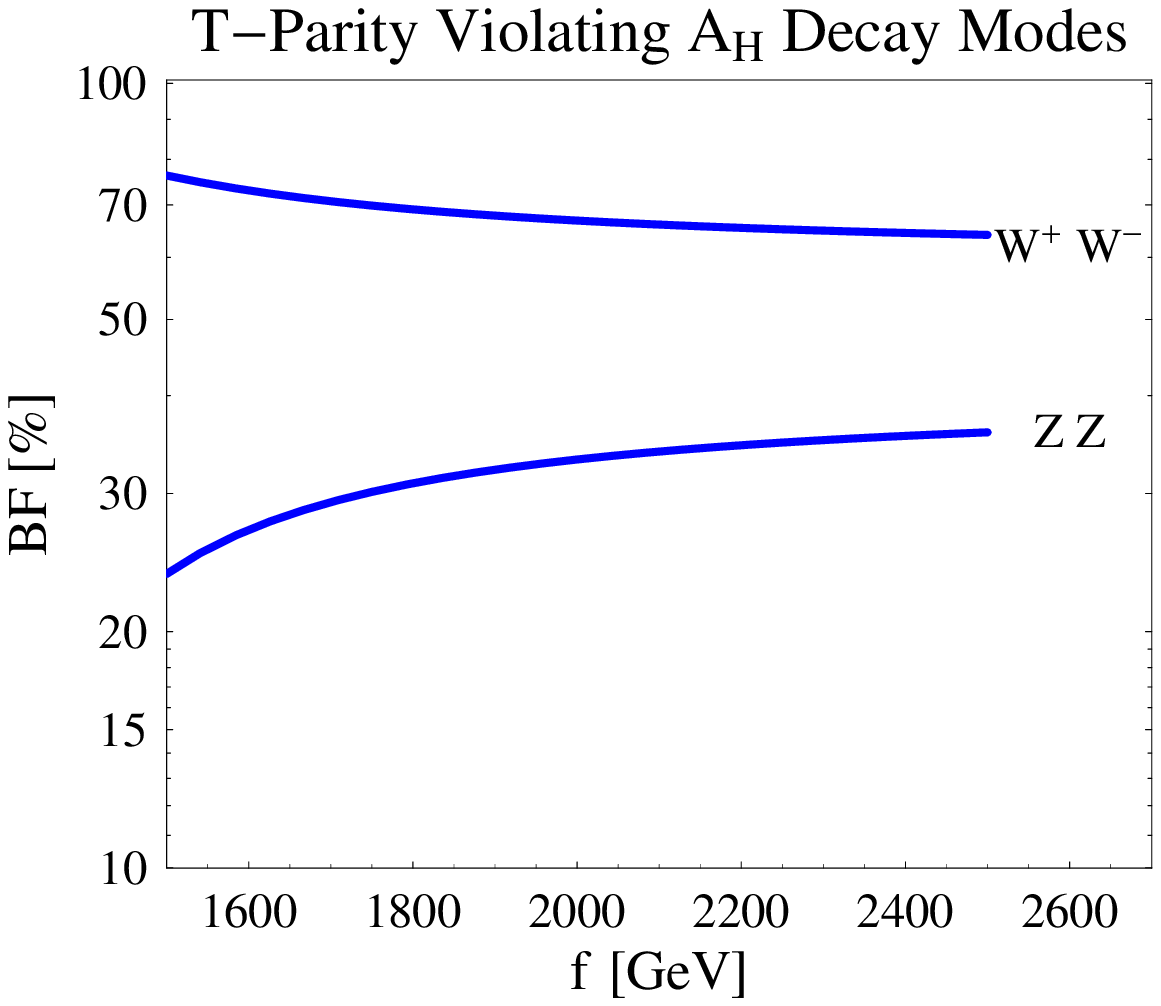}}
          \subfigure[Rest frame decay length of $A_H$]
            {\includegraphics[scale=0.60]{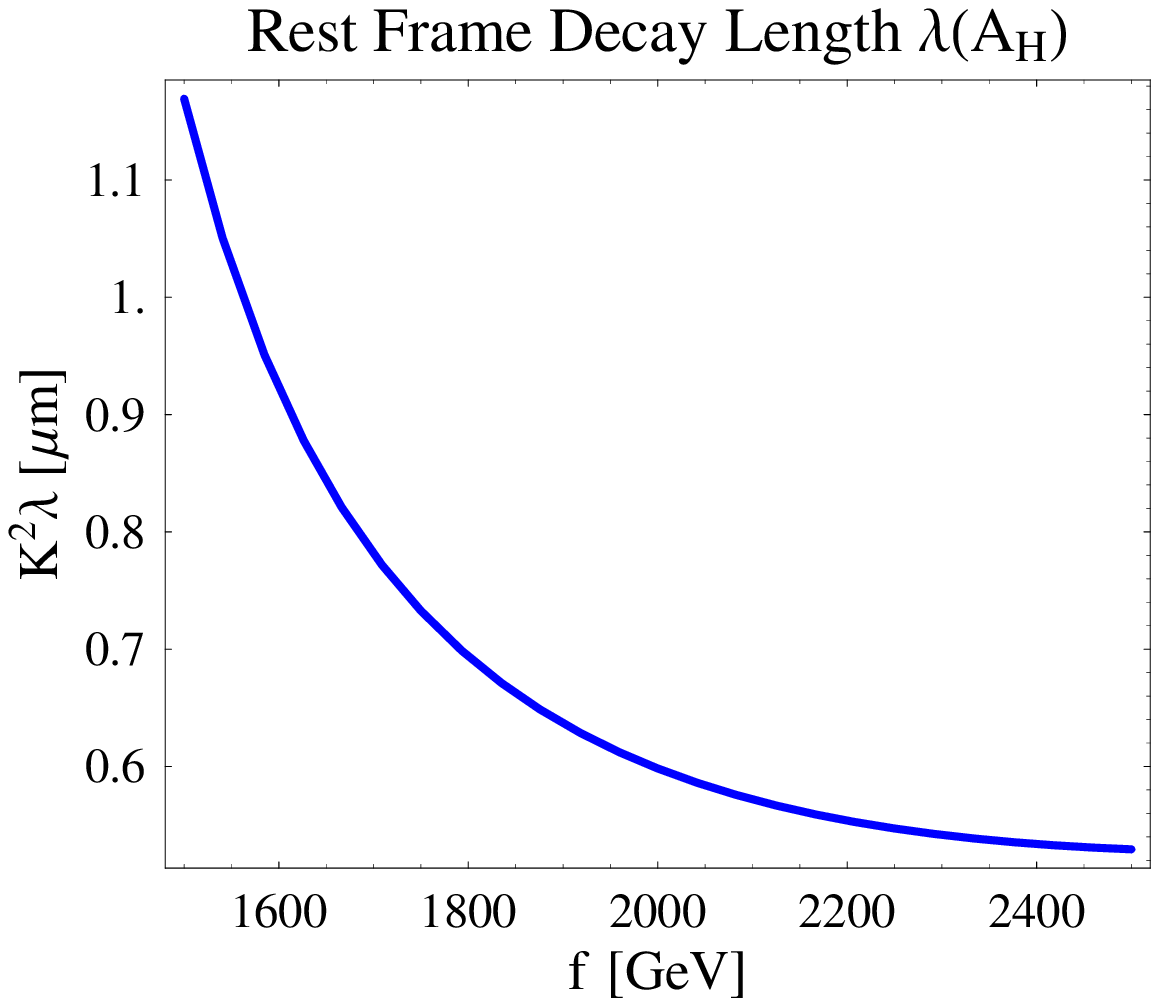}}\\
  \end{center}
  \caption{\footnotesize $A_H$ branching fractions and rest decay
length due to $\cal T$-parity violation.  K is the overall coefficient
in SU(5) Lagrangian, taken to be  $K= 1$ here. $A_H$ has a highly
suppressed decay width compared to the other LHT particles.}
\label{fig:TV}
\end{figure}

The leading order anomaly terms containing $B_HW\partial W$ and $B_HB\partial B$
cancel in the sum of WZW terms, and the remaining ${\cal
T}$-parity-violating terms have the forms $H^{\dagger}HB_HW\partial W$
and $H^{\dagger}HB_HB\partial B$.\cite{Hill:2007zv} The $B_H$ field is the $\cal T$-odd
partner of the SM axial $B$ field with parameters
\begin{equation}
m_{B_H} \simeq  {g' f}/{\sqrt{5}} \ ,
\quad  \tilde g=g'/\sqrt{5}  \ .
\end{equation}
The WZW term allows the $\cal T$-odd $B_H$ field to couple to 
$\cal T$-even SM gauge fields. The leading
relevant interaction is
\begin{equation}
{\cal L}_{\rm WZW} \supset  
-\frac{K\tilde{g}g^2_2N_{WZ} v_{SM}^2 }{48\sqrt{3}\pi^2 f^2} 
\epsilon^{\mu\nu\rho\sigma} {B_H}_{\mu} 
\left[ 
\sec ^2\theta_W  Z_{\sigma} \partial_{\nu}  Z_\rho 
+  (D^A_{\nu} W^+_{\rho}) {W}_\sigma^- 
+  (D^A_{\nu} W^-_{\rho}) {W}_\sigma^+ 
\right] 
\label{eq:psWZW}
\end{equation}
where $\theta_W$ is the electroweak mixing angle.
$K$ is an overall factor for the littlest $SU(5)/ SO(5)$ model. 
The WZW quantized integer $N_{WZ}$ is taken to be 3. 
The leading anomaly induced decays of $B_H$ in the LHT
model are $B_H\rightarrow ZZ$ and $B_H\rightarrow W^+W^-$.
Their partial widths are
\begin{equation}
\Gamma(B_H\to ZZ)={1\over 2\pi}
\left( K\tilde g^3 N_{WZ}\over 144\pi^2\right)^2 {m_Z^2\over m_{B_H}}\
\left(1-{4m_Z^2\over m_{B_H}^2}\right)^{5\over 2}
\label{eq:BZZ}
\end{equation}
\begin{equation}
\Gamma(B_H\to W^+W^-)={1\over \pi}
\left(K \tilde g^3 N_{WZ}\over 144\pi^2\right)^2 {m_W^2\over m_{B_H}}\
\left(1-{4m_W^2\over m_{B_H}^2}\right)^{5\over 2}
\label{eq:BWW}
\end{equation}
Details of the calculation are given in the Appendix.
$B_H$ is a combination of the $\cal T$-odd $A_H$ and $Z_H$ fields
\begin{equation}
B_H=A_H\ \cos{\theta_H}+  Z_H\ \sin{\theta_H} \ ,
\end{equation}
where $\theta_H$ is the mixing angle of the neutral heavy gauge bosons
at electroweak symmetry breaking, with its value given in Ref.\cite{Hubisz}
\[\sin{\theta_H}=\dfrac{5gg'}{4(5g^2-g'^2)}\dfrac{v_{SM}^2}{f^2}  \ .\]
Numerically the coefficient of the $Z_H$ term is negligible compared
to the coefficient of the $A_H$. {\it i.e.}, $B_H\approx A_H$. The
branching fractions of $A_H$ decay modes are shown versus $f$ in
Fig.\ref{fig:TV}.  The $A_H\rightarrow ZZh$, $W^{-}W^{+}h$ processes are
kinematically forbidden at natural $f$ values near 1 TeV.
In contrast $Z_H$ has many other decay senarios available will readily
 decay through dominant $\cal
T$-preserving modes discussed in the next section.

\begin{figure}[ht]
  \begin{center}
   \includegraphics[scale=0.65]{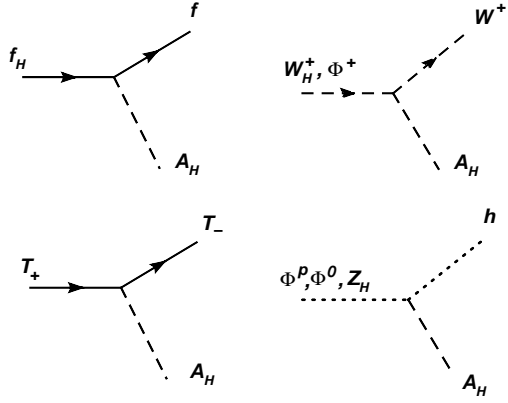}
  \end{center}
 \caption{\footnotesize Diagrams of the leading decay modes that
 produce $A_H$. $f$ stands for a fermion and $f_H$ for the heavy
 counterpart. Note that the $\cal T$-even $T_+$ decays into $A_H$ and
 $T_-$.}
\label{fig:LEAD}
\end{figure}

$A_H$ is not a viable dark matter candidate due to these $\cal
T$-violating decays. The total decay width of $A_H$ is found to be
$\sim10^{-1}$eV; the dependence of the $A_H$ rest decay length
($\lambda=c\hbar / \Gamma$) versus $f$ is plotted in the right panel of
Fig.\ref{fig:TV}.  The typical width of $\sim$ eV corresponds to a short track
of micrometers, which is practically an instantaneous decay.

$A_H$ is always a daughter particle of the decays of all other heavy
particles in LHT, as illustrated in Fig.\ref{fig:LEAD}. 
Thus the decays of $A_H$ greatly enhance the number
of final state gauge bosons, instead of contributing to missing energy
as expected in a strictly $\cal T$-parity conserving model.


\section*{Masses and Decay Widths of Heavy Particles in LHT}

At the LHC the new heavy particles in the Little Higgs model with $\cal T$-parity
(LHT) can be copiously
produced. For an enumeration of the different production channels see
\cite{Belyaev:2006jh}. Table \ref{tab:xsec} gives representative cross sections of
the leading heavy quark and gauge boson production processes at LHC.
\begin{table}[ht]
	\begin{center}
 	\begin{tabular}{|c|c|}
	\hline
	Final state&$\sigma$ [fb]\\
	\hline
	$q^+q^-$&5.2	\\
	$q^+q^+$&2.6  \\
	$T_-\bar{T}_-$&1.5 \\
	$q_-W_H^{+}+q_-W_H^{-}$&1.8\\	
	$q_-Z_H$&0.90\\
	$Z_HW_H^{+}\ +\ Z_HW_H^{-}$&1.6\\
	$W_H^{+}W_H^{-}$&1.0\\
	\hline
	\end{tabular}
	\end{center}
\caption{ Cross sections at the LHC of leading production processes
from $p\ p \rightarrow X X'$. We take $f = 1.5$ TeV, $\kappa = 1$ and
$m_h = 200$ GeV. In the left column $q^+ =
(u_-,c_-,\bar{d}_-,\bar{s}_-)$, $q^- = (\bar{u}_-,\bar{c}_-,d_-,s_-)$,
$q_- = (u_-,c_-,t_-,d_-,s_-,b_-)$ and the cross section is the sum of
contributions from all heavy quarks in the corresponding set.}
\label{tab:xsec}
\end{table}
%

\begin{figure}[ht]
  \begin{center}
   \includegraphics[scale=0.7]{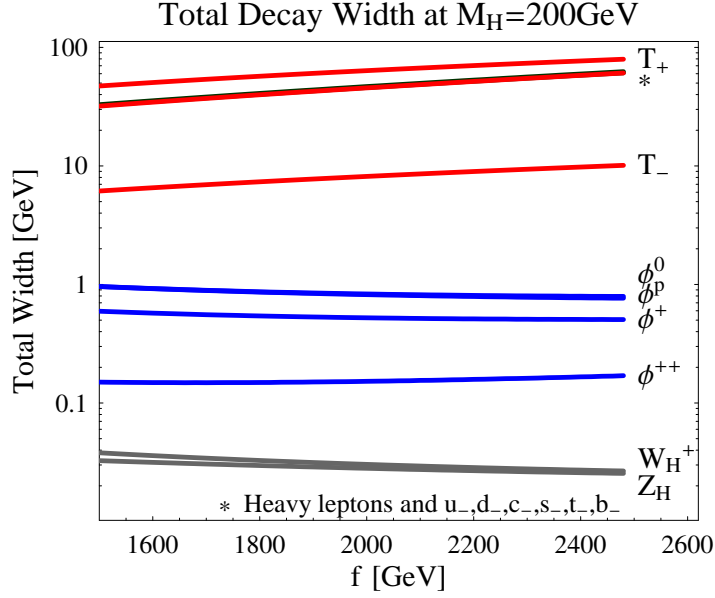}
  \end{center}
  \caption{\footnotesize 
Total widths of different parent heavy particles in LHT. The calculation assumes
$m_H$=200 GeV for heavy Higgs boson widths. In the figure the symbol $*$ denotes
$\cal T$-odd partners of the SM quarks and leptons.
}
  \label{fig:TOTW}
\end{figure}

Fig.\ref{fig:TOTW} gives the total decay widths of the massive quarks and gauge
bosons. Besides the dominant two-body decay modes, many three-body
decay channels may not be negligible for the reason that the
interactions of the longitudinal polarization of the gauge bosons are
enhanced and the bosonic couplings are large.  Due to the large mass
gap between the SM particles and the heavy partners, and among the
heavy particles themselves, three-body decays are usually accessible
(for instance $T_\pm $ at TeV mass) in strong contrast to the
restrictive three-body decay channels $t\to bWZ$\cite{Barger:1989ur}
and $t\to bWH^0$\cite{Barger:1987bv} of the top quark in SM.  As the
energy scale $f$ increases, the phase space of many three-body
channels opens up, and their branching fractions become experimentally
relevant at higher $f$ values. The $f$ dependence of significant
three-body decays is plotted in Fig.\ref{fig:bdec} and Fig.\ref{fig:fdec}.
The three-body channels can provide a good testing ground for the
detailed structure of the LHT interactions.

The particle table, Feynman rules, tables of parameters, and event
simulations of the LHT model have been coded in a public 
package CalcHEP LHT\cite{Belyaev:2006jh}
available at
{\tt http://hep.pa.msu.edu/LHT} 
for the phenomenology of the LHT model.  We make
use of this convenient tool to calculate various $\cal T$-parity
preserving multi-body decay channels.
In our analysis we fix the parameter $\kappa=1$, the SM Higgs mass
$m_h=$200 GeV, and take 1.5 $<f<$ 2.5 TeV as a compromise between
naturalness and the electroweak precision
constraints.\cite{Hubisz:2005tx} We do not include photons among the
daughter particles because of their suppressed coupling.

\section*{Multiple Body Decays of Heavy  Bosons}

The lightest $\cal T$-odd particle is the heavy photon $A_H$.  Exact
$\cal T$-parity conservation would require that normally $A_H$ be a
final decay product from any $\cal T$-odd heavy particle.  On the
other hand, when $\cal T$-parity is violated by the anomaly
interaction\cite{Hill:2007zv}
the $A_H$
will decay rapidly into $ZZ$ and $W^+W^-$.  The final products of the
$A_H$ decay can be detected, and the event kinematics can thereby be
fully reconstructed. 
The identification of three-particle decay modes becomes feasible.

Branching fractions of the leading decay channels are plotted versus
$f$ in Fig.\ref{fig:bdec}. We retain the channels with a fraction above 0.1\%. It
is interesting to see what new channels are available:

\begin{itemize}
\item[(i)] We start with the $\cal T$-odd neutral $Z_H$.  The two-body
decay mode $ Z_H \to A_H h$ dominates. A fermionic final state is not
kinematically allowed.  However, the three-boson phase space is open
and there is a substantial branching fraction at the level of 10\%
for $Z_H \to W^+ W^- A_H$.

\item[(ii)] $W_H^+ \to W^+ A_H$ is the dominating two-body
mode in $W_H^+$ decay. Similar to $Z_H$ decay, $W_H^+$ is less massive than
heavy fermions and any fermionic final state is kinematically 
disallowed. The heavy mass $M_{W_H}$ allows
both $W^+A_H h$ and $W^+A_H Z$ decays at the level of a few percent.

\item[(iii)] The decay of the singly charged $\phi^+$ is mainly
dominated by the two-body mode $\phi^+ \to W^+ A_H$. Three-body
channels $W^+A_H h$ and $W^+A_HZ$ also give significant contribution to
the total width.

\item[(iv)] The neutral component $\phi$ of the triplet scalar boson
is a complex field. It is decomposed into the real part (a scalar
$\phi^0$) and the imaginary part (a pseudoscalar $\phi^p$) of
approximately equal masses. The different spatial parities imply
different decay modes.
$$ \begin{array}{cc}
\phi^0 \to Z A_H  \ ,& \phi^0 \to Zh A_H \ ;\\
\phi^p \to h A_H  \ ,& \phi^p \to ZZ A_H \ .    \end{array}  $$
Note the role swap $Z\leftrightarrow h$ when $\phi^0\leftrightarrow
\phi^p$. Both $\phi^0$ and $\phi^p$ decay to $W^+W^- A_H$ as well.

\item[(v)] The doubly charged scalar boson $\phi^{++}$ cannot decay
into $\phi^+ W^+$ because of the common $\phi^{++}, \phi^+$ mass. At
low Higgs mass $\sim$120 GeV there are no two-body decays of
$\phi^{++}$; however $\phi^{++} \to W_H^{+} W^+$ emerges at higher
Higgs mass when $\phi$ becomes much more massive than heavy weak gauge
bosons. The virtual process $\phi^+_{\rm virtual} \to W^+ A_H$ gives
the overall leading three-body decay $\phi^{++} \to W^+ W^+ A_H$.
The four-body decay channels ($\phi^{++} \to W^+W^+A_H h$ or
$W^+W^+A_HZ$) are smaller but can still be of comparable size to the
two/three body-decay modes because of the limited width of leading decay
modes.
\end{itemize}

\section*{Multiple Body Decays of Heavy Quarks}

There are many $\cal T$-odd fermions $f_-$ of different flavors
$(u_-,d_-,c_-,s_-,t_-,b_-,T_-)$ with TeV scale masses.  $\cal
T$-parity demands at least one heavy boson in the final state. 

The general pattern of decay channels according to descending
branching fractions are\\
\hspace*{1mm} $\ \ f_{-} \rightarrow\ \ W_H + f' \ , 
              \ \ \ \ Z_H+ f\ ,\ \ \ \ A_H+ f\ \ $    or,\\
\hspace*{1mm} $\ \ f_{-} \rightarrow\ \ W+W_H+ f \ , 
\ \ \ \ W+Z_H+ f' \ , \ \ \ W_H+ Z+ f' \ , \ \ \ W+ A_H+ f' \ ,
 \ \ \ Z+ Z_H+ f   \ $\\

\noindent
where $f,f'$ are the SM doublet.  The relevant Feynman diagrams are
shown in Fig.~5.

\begin{figure}[ht]
\begin{center}
 \includegraphics[scale=0.7]{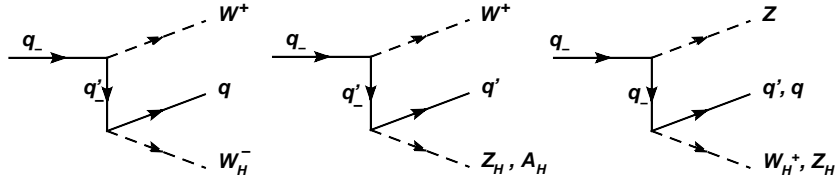}
\caption
{\footnotesize Feynman diagrams for the three-body decay processes of
a ${\cal T}$-odd fermion.  $q,q'$ refer to different flavors of a SM
doublet.}
\end{center}
\label{fig:FEYN}
\end{figure}

The $T_+$ quark of even $\cal T$-parity decays readily through $t-T_+$
mixing into $W^+b$, $Zt$, $ht$ as dominating two-body modes.  The
three-body mode $Zht$ occurs as a rare process, in analog to the rare
decay modes of the fourth-generation quark as studied in
Ref.\cite{Barger:1989ur}. In addition, being $\cal T$-even $T_+$ has a
rare mode of decaying into two heavy photons.

Note that some three-body channels that can cascade from the primary
on-shell two-body decay modes are not shown in our plots, mainly
because their rates depend very much on the mass cuts in separating
out the resonance components; for example, $u_-\to d (W^+ A_H) $ where
$(W^+ A_H)$ can be the resonance of $W^+_H$. These channels with
intermediate resonance would nevertheless be very important in
determining the resonance masses.

\section*{Detection}

Little Higgs phenomenology at the LHC has been investigated recently
in a number of studies\cite{PHENO} but not for the situation where $A_H$
decays. The key feature is that the new heavy quarks can
also be produced in proton-proton collisions, either from gluon fusion
or quark interactions.  Most $\cal T$-odd quarks are pair produced with
the exception that the $\cal T$-even $T_+$ particle can be
produced along with a SM quark.

The produced heavy quarks decay quickly into less massive SM particles
and $\cal T$-odd bosons that subsequently decay into SM counterparts
and heavy photons. The dominant two-body decay channels $\ f_{-}
\rightarrow\ W_H^{\pm}+ f'\ , \ Z_H+ f \ ,\ A_H+ f\ \ $ transform
each heavy quark into a SM quark that may form a jet and one $\cal T$-odd
gauge boson.  $W_H^{\pm}$ and $Z_H$ decay into SM gauge/Higgs bosons and
$A_H$. $A_H$ decays through $\cal T$-violating WZW interation 
into $ZZ$, $W^+W^-$, resulting in an overall $\cal
T$-even final state.

As shown in Fig.\ref{fig:FEYN}, 
the leading three-body decay processes of $\cal T$-odd quarks will either add a $W^\pm$
or $Z$ boson to
the daughter particles, while $T_{\pm}$ gives an additional $Z$ boson
in the final state. In the bosonic sector, the heavy gauge
bosons have significant decay rates to three-body final states.

The contributions of the three-body decays visibly depend on the
energy scale $f$. As $f$ increases to higher $f$ values, the three-body phase space opens
up faster than the two-body phase space, and the three-body branching
fractions steadily increase as the mass gap widens between the heavy
particles and SM particles.

\section*{Conclusion} 

The Little Higgs model with $\cal T$-parity (LHT) is an interesting
extension of the Little Higgs framework. It alleviates the tension of
the ``little hierarchy" problem and it is also a phenomenologically
rich model, giving rise to testable new physics at the TeV scale.

In the LHT model one can expect that the LHC will produce a large
amount of heavy quarks beyond the SM via the strong interaction, and
also substantial numbers of new heavy leptons and new heavy gauge bosons by
Drell-Yan-like processes. Their decay patterns can go beyond the usual
dominant two-body modes and include contributions from various
measurable three-body modes.

Since $\cal T$-parity is broken by anomaly, the lightest $\cal T$-odd
particle $A_H$ will decay into detectable $ZZ$ or $W^+W^-$.  As $A_H$
appears as a decay product of all LHT processes, the $\cal
T$-parity-violating decays allow reconstruction of the full event
configurations and thereby comprehensive physics tests of the Little Higgs model
at the LHC.

We have studied the multi-body decays of the heavy particles in the
LHT model that can be produced at the LHC. Detailed analyses of these
multi-body channels may be useful in revealing the new symmetry and
its interactions at the TeV scale.

\section*{Acknowledgements}

We thank Christopher T.~Hill and Richard~J.~Hill for very helpful
discussions. This research was supported in part by the U.S.\
Department of Energy under Grant Nos.\ DE-FG02-95ER40896 and
DE-FG02-84ER40173, and in part by the Wisconsin Alumni Research
Foundation.

\section*{Appendix}

The decay amplitude 
$ B_H (\varepsilon') \to Z(k_1,\varepsilon_1) +  Z(k_2,\varepsilon_2)$
can be derived from Eq.~(\ref{eq:psWZW}) 
\begin{equation}
{\cal M}=-{K\tilde g^3 N_{WZ} \  \ m_Z^2
\over 12\sqrt{3}\ \pi^2 M_{\tilde B}^2} L \ ,\quad
L=\epsilon^{\mu\nu\rho\sigma} 
\varepsilon'_\mu (k_1-k_2)_\nu  (\varepsilon_1)_\rho
                                (\varepsilon_2)_\sigma
 \ ,
\end{equation}
where the Levi--Civita symbol is contracted with vectors.  The
momentum term $k_1-k_2$ comes from two ways of contracting the $Z$
field. It antisymmetrizes the momentum part and the Levi--Civita
antisymmetrizes the polarization part.  The combined product is
overall symmetric as expected for the boson decay. We choose
$\varepsilon'$ in the rest frame of $B_H$ along the $\bf z$ direction,
and ${\bf k}_1=-{\bf k}_2={\bf k}=|{\bf k}| (\sin\theta {\bf x}
+\cos\theta{\bf z})$. Notice that transverse-transverse (TT) modes
vanish, as well as the  longitudinal-longitudinal (LL) mode. The
only surviving modes are LT or TL. The relevant vectors in the LT mode are
$$ \begin{array}{rlrrrr}
      \varepsilon'=(&0, &0,          & 0, &1)& \\
     k_1-k_2  =(&0, &\sin\theta, & 0, &\cos\theta)&2|{\bf k}| \\
     \varepsilon_1(L)=(&|{\bf k}|, &E_Z \sin\theta,& 0, &E_Z\cos\theta)&/m_Z \\
     \varepsilon_2(T)=(&0, &  0,       & 1, &0)& \end{array}
$$
\noindent
The Levi--Civita symbol becomes the determinant of the above arrays,
$L=2|{\bf k}|^2\sin\theta /m_Z$.
$$ \sum_{\rm final}|{\cal M}|^2
=\left({K \tilde g^3 N_{WZ}\over 144\pi^2}\right)^2
12 {m_Z^4\over m_{B_H}^4}\ 
{m_{B_H}^4\over m_Z^2}
\left(1-{4m_Z^2\over m_{B_H}^2}\right)^2\sin^2\theta \times 2
$$
The last factor two counts both  LT and TL modes.  Thus we obtain
$$ \Gamma(B_H\to ZZ) ={1\over 2m_{B_H}}{1\over
  8\pi}\left(1-{4m_Z^2\over m_{B_H}^2}\right)^{1\over 2}
\sum_{\rm final}|{\cal M}|^2\left({d\Omega\over 4\pi}\right) {1\over 2!}
$$
The factor $1\over2!$ in the above decay width comes from the
combinatorics of the two identical $Z$ bosons. 
After some algebra, we derive the final expressions
Eqs.~(\ref{eq:BZZ}, \ref{eq:BWW}) of $\Gamma(B_H\to ZZ)$, and the
similar one $\Gamma(B_H\to W^+W^-)$.
The threshold dependence agrees with that in Ref.~\cite{Chang:1988fq}.
Note that the overall factor of 2 difference between $WW$ and $ZZ$ comes from
identical particle effect.


\begin{figure}[ht]
  \begin{center}
  \includegraphics[scale=0.63]{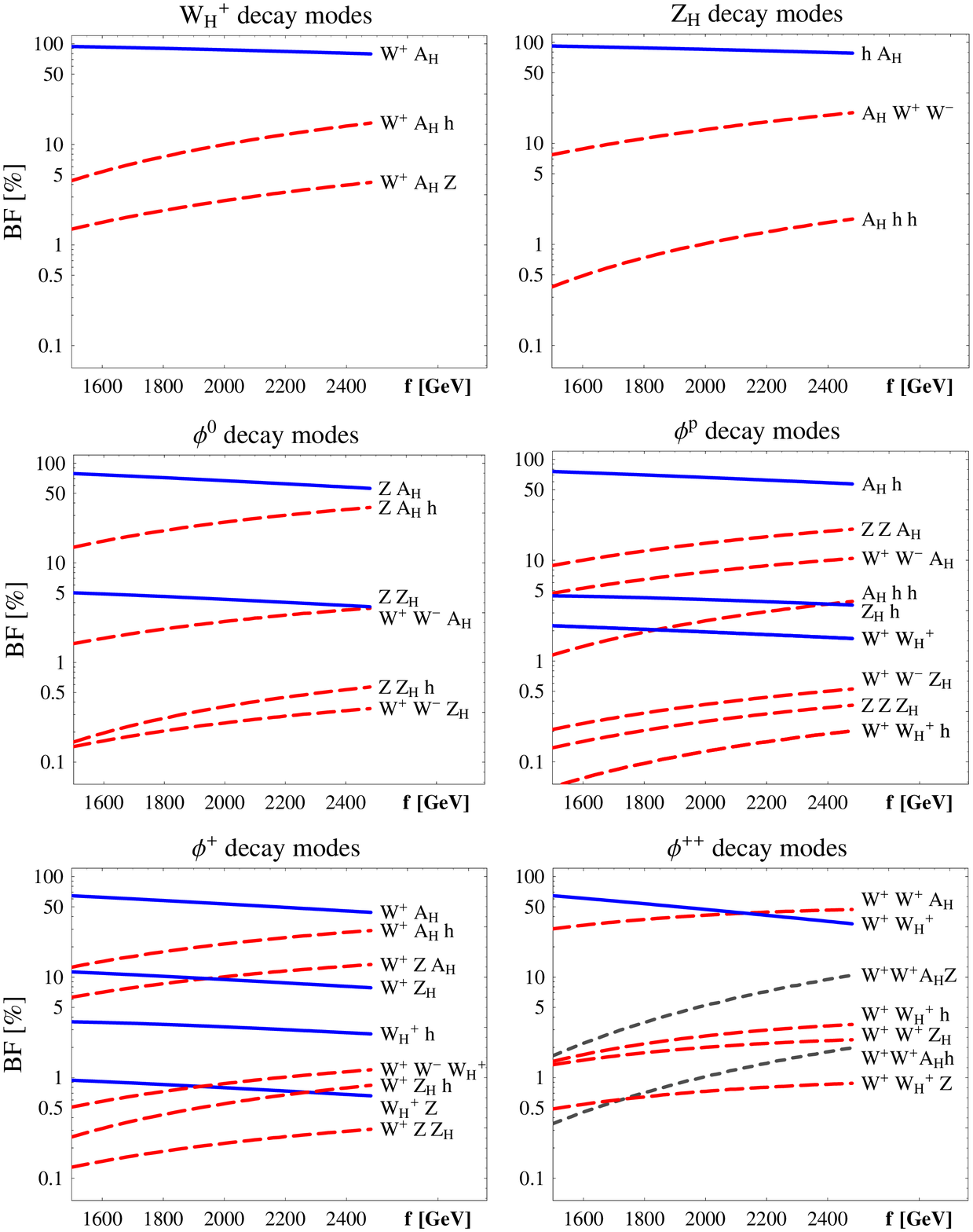}
  \end{center}
  \caption{\footnotesize Branching fractions of heavy boson decay modes in LH are
plotted versus the global symmetry breaking scale $f$ (GeV). Solid lines (Blue) and
long dashed lines (Red) show two- and three-body channels, respectively. Due to
the limited width of the two-body mode of $\phi^{++}$, the leading
four-body modes also reach high branching franctions, as shown in short
dashed lines (Grey).}
  \label{fig:bdec}
\end{figure}

\begin{figure}[ht]
  \begin{center}
  \includegraphics[scale=0.63]{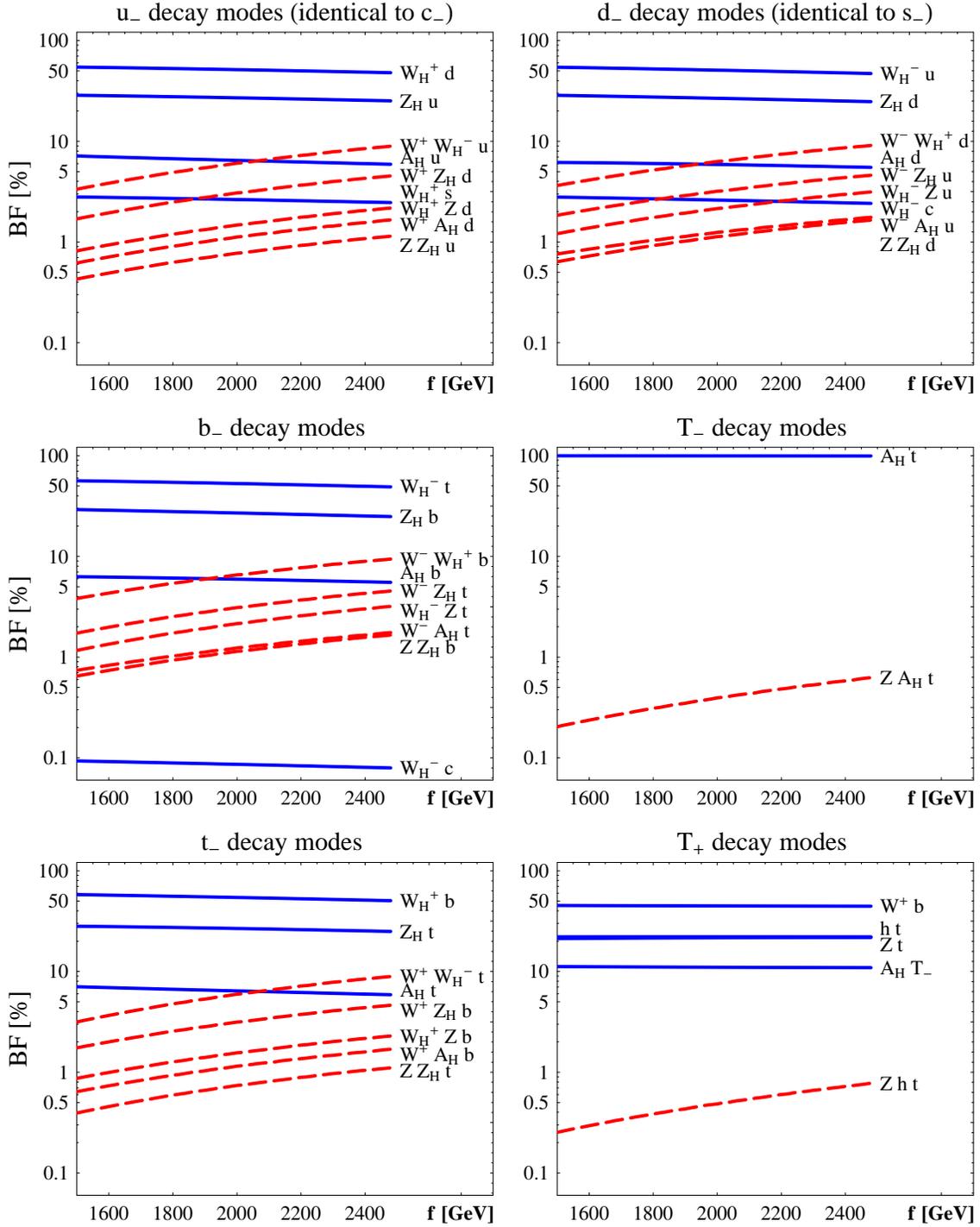}
  \end{center}
  \caption{\footnotesize 
Branching fractions of heavy fermions in LH are plotted versus $f$. 
Cases of parent particles
$u_-$ (or similarly $c_-$), $d_-$ (or $s_-$), $b_-$, $T_+$, $t_-$, 
are shown in the composite graphs.  
The branching fractions are calculated at $m_h$=200 GeV
when Higgs bosons are involved.}
 \label{fig:fdec}
\end{figure}

\end{document}